\documentclass[%
 amsmath,amssymb,
 aps,
prfluids
]{revtex4-2}

\usepackage{graphicx}
\usepackage{dcolumn}
\usepackage{bm}
\usepackage{hyperref}

\usepackage[
text={7in,10in},centering,
]{geometry}

\usepackage{lineno}

\newcommand{\etal}{\emph{et al.\ }}
\newcommand{\ecoli}{\emph{E. coli }}
\begin{document}

\title{Effects of shear thinning viscosity and viscoelastic stresses on flagellated bacteria motility}


\author{Zijie Qu}
 \altaffiliation[Current address: ]{California Institute of Technology, The Division of Biology and Biological Engineering, 1200 East California Boulevard, Pasadena, California, 91125 USA.\\
 Email: zijiequ@caltech.edu}
\author{Kenneth S. Breuer}%
\affiliation{%
 Brown University, School of Engineering, 184 Hope St, Providence, RI 02912 USA
}%

\date{\today}

\begin{abstract}
The behavior of flagellated bacteria swimming in non-Newtonian media remains an area with contradictory and conflicting results. We report on the behavior of wild-type and smooth-swimming \emph{E. coli} in Newtonian, shear thinning and viscoelastic media, measuring their trajectories and swimming speed using a three dimensional real-time tracking microscope. We conclude that the speed enhancement in Methocel solution at higher concentration is due to shear-thinning and an analytical model is used to support our experimental result. We argue that shear-induced normal stresses reduce the wobbling behavior during cell swimming but do not significantly affect swimming speed.  However the normal stresses play an important role in decreasing the flagellar bundling time which changes the swimming speed distribution. A dimensionless number, the ``Strangulation number'' ($Str$) is proposed and used to characterize this effect.
\end{abstract}
\maketitle

\section{Introduction}
\label{sec:Introduction}
Many microorganisms live in various aquatic environments and propel themselves by rotating effectively rigid helical flagella \cite{berg1972chemotaxis} or undulating flexible cilia \cite{lighthill1976flagellar,fauci1995sperm}. Most of these cells live in biological fluids, such as mucus, which can exhibit complex non-Newtonian properties \cite{suarez2006sperm}, and one issue that has received recent attention to resolve is how and why non-Newtonian effects change cell swimming characteristics.  

Experimental studies of swimming in viscous and non-Newtonian fluids have reported different, apparently contradictory results.  Berg and Turner \cite{berg1979movement}  measured the rotational speed of wild-type tethered \ecoli and discovered a non-monotonic change in rotational speed as a function of viscosity. In addition, the rotational speed was different in solutions of Ficoll, a branched polymer with Newtonian characteristics,  and Methocel, a long-chain, unbranched polymer with viscoelastic and shear-thinning properties.  These differences were evident even when the media exhibited the same bulk viscosity. Berg and Turner concluded the difference was due to the interactions between flagellar filaments and the quasi-rigid polymer networks.  

Two non-Newtonian effects are likely to influence the mechanics of flagellar swimming:  shear thinning and the presence of normal stresses.  Shear thinning, in which the high rotation rate of the flagella decreases the effective viscosity, has been proposed several times including in the original Berg and Turner experiments \cite{berg1979movement} as well as more recently by Martinez \etal \cite{martinez2014flagellated}. Theoretical models based on Resistive Force theory (RFT) and a two-viscosity model applied to the cell body and flagellum respectively have also been presented to support these data \cite{martinez2014flagellated}. In addition, Gomez \etal \cite{gomez2017helical} argued that although the swimming speed is enhanced in a shear thinning fluid, it should be explained by a result of viscosity gradient rather than by a two-viscosity model assuming a constant speed motor, which differs from the motor behavior of bacteria \emph{E. Coli} \cite{chen2000torque}, in power law fluid with varying indices.  The results appear to be dependent on the details of the motion, for example, as demonstrated by  Montenegro-Johnson \emph{et al.} \cite{montenegro2013physics}, who found that the swimming speed of an idealized two-dimensional undulating sheet in a shear-thinning fluid could be either enhanced or hindered, depending on the details of the flagellar kinematics. 

Other studies have argued that non-Newtonian normal stresses might also be responsible for the observed speed up in polymeric swimming. Patteson \emph{et al.} \cite{patteson2015running} tracked wild-type \ecoli in viscoelastic polymer solutions and found that the cells swam faster and in a straighter path as the polymer concentration increased. They argued that normal stresses introduced by the elastic properties of the fluid reduced the cell body wobble, explaining both of these observations.  However, the effect of cell precession (wobble) on swimming speed is not at all clear.  While Darnton and Patteson both observed an anticorrelation between wobble and swimming speed \cite{darnton2007torque,patteson2015running}, Liu \emph{et al.} \cite{liu2014helical} argued, based on measurements of wild-type \textit{C. crescentus} and a modified RFT, that cell precession generates thrust and increases swimming speed.  However, unlike \ecoli, \textit{C. crescentus} is a uni-flagellated bacteria with a crescent-shaped cell body  which swims using a ``run-reverse-flick'' strategy, and these differences, particularly the cell geometry, may lead to differences in the role of wobble on swimming speed.  

Non-Newtonian stresses are also known to both increase and decrease the propulsive speed of a single rotating helical filament, depending on the helix geometry and the Deborah number, $De$, where the Deborah number is the product of the rotational speed, $\omega$ and the elastic relaxation time, $\tau$.  Experiments using a model helical filament in a non-shear-thinning viscoelastic (Boger) fluid \cite{liu2011force} demonstrated increased swimming speeds, while computations conducted over a wider range of geometric and flow parameters \cite{spagnolie2013locomotion} showed both accelerated and retarded speeds due to the non-Newtonian fluid properties. 

In live cell experiments, two complicating factors have arisen that make comparisons between different experiments and theory challenging.  Firstly  most of the experiments have observed the behavior of wild-type \ecoli cells \cite{berg1979movement,martinez2014flagellated,patteson2015running}, which exhibit a ``run and tumble'' style of motility \cite{berg1972chemotaxis}. This is problematic because, as mentioned above, the flagellar bundling process of a wild-type swimmer is quite sensitive to changes in viscosity \cite{qu2018changes}, and this affects the average run speed as well as the distribution of speeds observed over many run-tumble cycles. Although it has not yet been studied, the bundling process is also likely to be affected by non-Newtonian fluid affects.

A second factor that makes experimental observations difficult to compare stems from the composition of the base media.  Recent studies \cite{martinez2014flagellated,qu2018changes} have demonstrated that short chain fragments in the polymer solutions have strong effect on the activity level of the cells, increasing the average run speed.  A consensus has arisen that only by dializing the solutions to remove the 
polymer fragments are reliable comparisons between swimming in baseline (non-polymer) and enhanced (polymer) media achievable.

Lastly, we note differences in defining the speed enhancement. Several reports have compared the speed in a non-Newtonian solution to the speed in a Newtonian solution with the identical shear viscosity
\cite{gomez2017helical,liu2011force,spagnolie2013locomotion} and have reported an increase in speed. However, it is known that the swimming speed of a cell (with constant torque motor) decreases with increased shear viscosity \cite{magariyama2002mathematical} which could lead to an overall decreasing trend if absolute swimming speed is studied as a function of viscosity in non-Newtonian solution. However, such enhancements have been observed from several experimental studies \cite{martinez2014flagellated,qu2018changes,patteson2015running}.

For all these reasons, the results from theoretical, numerical and experimental perspectives remain controversial and there is still no clear understanding of the relative importance of different non-Newtonian effects on cell swimming behavior. With this study, we hope to resolve these issues and we present results on the swimming behavior of both wild-type and smooth-swimming \emph{E. coli} in a variety of fluid media: a Newtonian motility buffer with a range of viscosities (Ficoll 400) and different concentrations of a viscoelastic, shear-thinning medium (Methocel). In all cases, the fluid media are dialized to remove any polymer fragments. Smooth swimmers do not tumble, thus providing a means to measure the pure swimming effectiveness of cells.  Data on the behavior of both smooth-swimming and wild-type cells thus provides us a means to separate the swimming mechanics from the bundling mechanics. Our experimental technique (Materials and Methods, Sec. A - D) tracks individual cells over long periods of time using a three-dimensional tracking microscrope which accounts for cell-to-cell variations (due to  variations in cell geometry, etc). 

In the following section (Sec.~\ref{sec:Results}), we present our measurements of the swimming speed and speed distribution of both wild-type and smooth-swimming cells in the different fluid media.  In Sec.~\ref{sec:Discussion} we then use a range of  arguments and analyses to tease apart the roles of shear thinning and non-Newtonian normal stresses on swimming speed, cell-body wobble and flagellar bundling.

\section{Results}
\label{sec:Results}

\subsection{Swimming behavior of smooth swimmers in both Newtonian and non-Newtonian solutions}
\begin{figure}
\centering
\includegraphics[width=8cm]{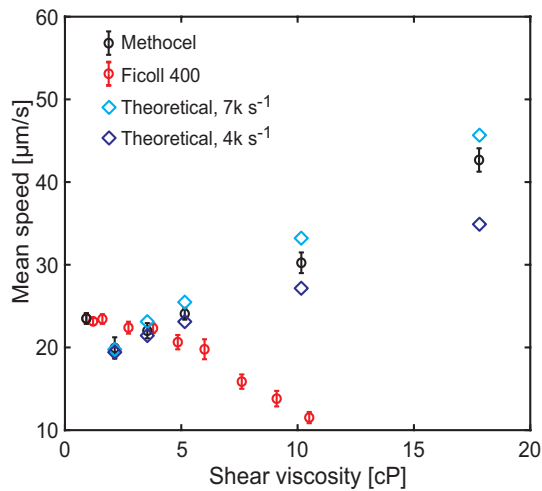}
\caption{Mean swimming speed of smooth swimmers in Ficoll (red) and Methocel solutions (black). The viscosity is the shear viscosity measured at $200$ $s^{-1}$. The swimming speed decreases with increased viscosity in Ficoll solutions, but increases with increased viscosity in Methocel solutions. Calculated swimming speeds, using a  shear-dependent RFT model \cite{martinez2014flagellated} are plotted using light and dark blue markers, for flagellar shear rates of $\dot{\gamma}_{f}$ = $7000$ $s^{-1}$ and $4000$ $s^{-1}$ respectively.}
\label{fig:fig1_speed}
\end{figure}

\begin{figure}
\centering
\includegraphics[width=8cm]{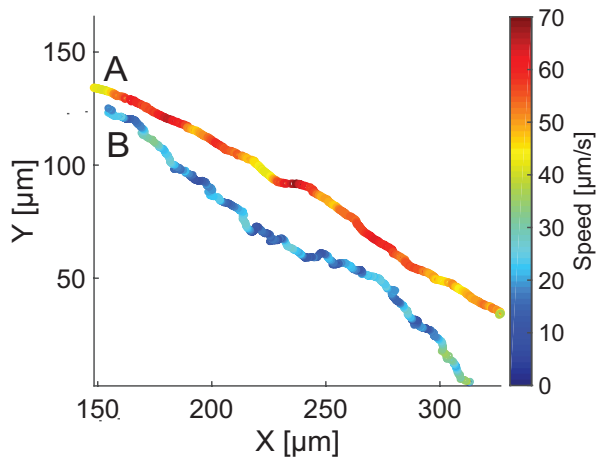}
\caption{The 2D projection of the swimming trajectories of two individual swimmers. A. In 0.500\% Methocel solution. B. In motility buffer. It is clearly seen that the trajectory in the Methocel solution is slower and smoother than the trajectory in the motility buffer.}
\label{fig:fig2_Traject}
\end{figure}

Using  three-dimensional real-time tracking of multiple cells, the behavior of smooth swimmers was observed in Ficoll and Methocel solutions of varying concentration. The average swimming speeds from at least $25$ individuals at each experimental condition are shown in Fig.~\ref{fig:fig1_speed}. Sample trajectories of two typical swimmers, one in motility buffer (Newtonian, viscosity $0.98$ cP) and one in Methocel solution (viscoelastic, bulk viscosity $17.80$ cP), are shown in Fig.~\ref{fig:fig2_Traject}. 

\subsection{Swimming behavior of wild-type cells in non-Newtonian solutions}
The swimming behaviors of wild-type \ecoli in Methocel solutions at various concentrations were also observed and analyzed. As previously suggested by Qu \emph{et al.} \cite{qu2018changes}, the skewness of an individual cell's swimming speed distribution provides a good measure of the swimming behavior and the relative amounts of time spent during run and tumble phases. The tumble behavior was defined by a sudden change in orientation and can be measured experimentally from the swimming trajectory. With increased viscosity, \ecoli spends an extended time recovering from tumble to run due to an elongated bundling process. The speed distribution of a wild type swimmer in buffer solution (low viscosity) is highly asymmetric because it spends most of the time running (high speed) and short time tumbling (low speed). The skewness measures how asymmetric a distribution is and is zero for a symmetric distribution. The value of wild type \ecoli speed distribution is more negative with shorter bundling time and becomes close to zero with longer bundling time as a result of increased viscosity. Moreover, a characteristic run speed can be estimated using the individual speed skewness and mean. It is noted by Qu \emph{et al.} \cite{qu2018changes} that at a given characteristic run speed, the mean speed of wild-type \ecoli cells is proportional to the skewness of the speed distribution using statistical simulation. The characteristic run speed is difficult to measure experimentally, especially when the solution viscosity is high that run behavior at full speed is less likely maintained.

\begin{figure}
\centering
\includegraphics[width=16cm]{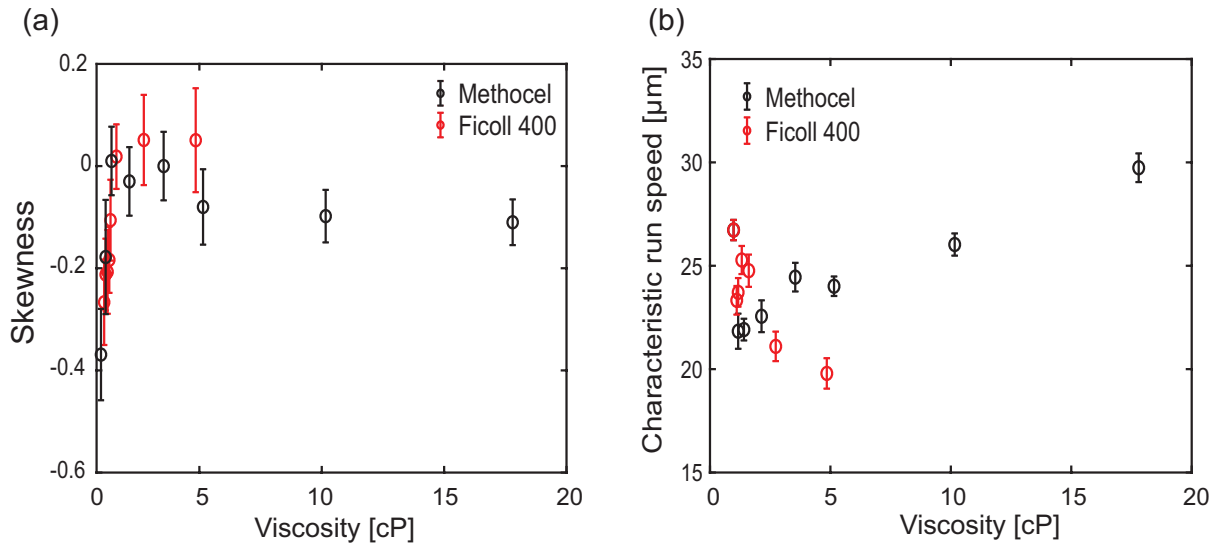}
\caption{(a). Averaged skewness of swimming speed distribution of wild-type cell in Methocel (black) and Ficoll (red) solutions at various viscosities. The skewness increases monotonically in Ficoll solution with increased viscosity. In Methocel solutions, the averaged skewness increases from negative to $0$ and then starts to decay.  (b). Averaged characteristic run speed of wild-type cells in Methocel (black) and Ficoll solutions (red) at various viscosities. The speed decreases in Ficoll solutions with increased viscosity, but increases with increased viscosity in Methocel solutions. (Ficoll results reproduced from Qu \emph{et al.} \cite{qu2018changes}).}
\label{fig:fig3_WT}
\end{figure}

Here, the skewness averaged  over all individuals being tracked as a function of viscosity is shown in Fig.~\ref{fig:fig3_WT}(a) while the characteristic run speed as a function of shear viscosity, calculated using the analysis described by Qu \etal \cite{qu2018changes},  is shown  in Fig.~\ref{fig:fig3_WT}(b). 

\section{Discussion}
\label{sec:Discussion}
It is observed that the average swimming speeds (averaged characteristic run speed) of both wild-type and smooth-swimming cells are enhanced significantly in Methocel solutions with increased shear viscosity (Fig.~\ref{fig:fig1_speed} black markers and Fig.~\ref{fig:fig3_WT} (a)), which is in sharp contrast to the decreasing trend of mean swimming speed in Newtonian solutions (Fig.~\ref{fig:fig1_speed}, red markers). This phenomenon has been previously observed \cite{patteson2015running,liu2011force,martinez2014flagellated}, but has been explained using different reasons including the shear-induced normal stress which reduces cell wobble \cite{patteson2015running}, viscoelastic stresses  \cite{liu2011force} and shear thinning of the polymer solutions \cite{martinez2014flagellated,gomez2017helical}. To understand different effects on the swimming speed, we first focus our analysis on smooth-swimming cells, since this isolates the swimming mechanics from any effects associated with flagellar bundle formation and breakup.

\subsection{Flagellar motor torque-speed behavior}
We start with characterization of the smooth swimmer flagellar motor behavior. As shown in Fig.~\ref{fig:fig1_speed} (red markers), the mean swimming speeds in (Newtonian) Ficoll solutions decrease as the solution viscosity rises and, although the decline is monotonic throughout the range of viscosities tested, there is an increase in the rate at which the speed decreases for $\mu > \sim 5$cP.  At the higher concentrations the speed decreases as $1/\mu$, suggesting that, in this regime the torque of the motor is constant. This has been previously observed experimentally \cite{chen2000torque,qu2018changes} and modelled analytically \cite{qu2018changes,magariyama2002mathematical}. In contrast, the swimming speed trend at lower viscosity implies that the torque of the motor is increasing with respect to its rotational speed \cite{chen2000torque,qu2018changes}.

\begin{figure}
\centering
\includegraphics[width=8cm]{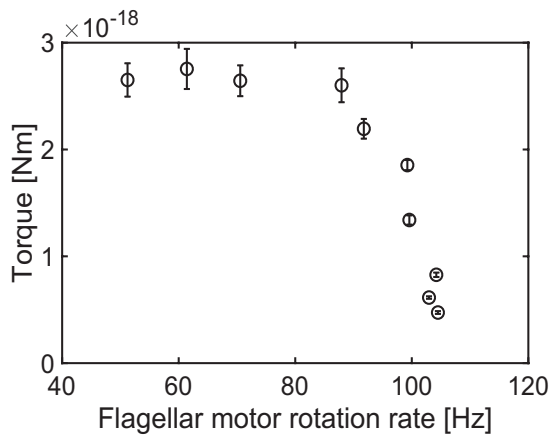}
\caption{Smooth swimmer flagellar motor torque behavior calculated using Resistive Force Theory.  The knee speed of the motor is about $100$ Hz which is a bit lower than that found in wild-type cells \cite{qu2018changes,chen2000torque}.}
\label{fig:fig4_Motor}
\end{figure}

Since we remain in the low $Re$ number ($Re \sim 10^{-4}$) regime \cite{man2017bundling,lauga2009hydrodynamics}, Resistive Force Theory (RFT) for the cell and helical bundle \cite{magariyama2002mathematical,qu2018changes} are used to estimate the torque-speed characteristics of the motor (Fig.~\ref{fig:fig4_Motor}). 
The geometry of the cell is included in Table SI1. The behavior is consistent with previous measurements \cite{darnton2007torque,qu2018changes,chen2000torque,reid2006maximum}, although the ``knee speed'' of the motor is a little slower and the stall torque a little larger than those found in previous observations of wild-type cells. 
With this reassurance that the cells studied are ``typical'' we address two hypotheses to explain the increased swimming speed observed: (i) the reduction in the cell wobble, or precession and (ii) shear thinning behavior of the Methocel medium.

\subsection{Shear-induced normal stress reduces wobbling effect}
\begin{figure}
\centering
\includegraphics[width=8cm]{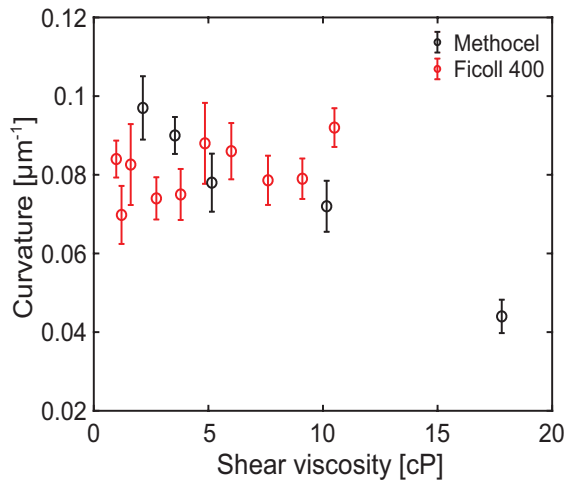}
\caption{Averaged local curvature of all swimming trajectories at different viscosities. Red markers show the results in Ficoll solutions and black markers are the results in Methocel solutions.}
\label{fig:fig5_Curvature}
\end{figure}

Patteson \emph{et al.} \cite{patteson2015running} measured the averaged wobbling angle and discovered that it decreased with respect to increases in polymer concentrations (viscosity). They also qualitatively demonstrated that the swimming trajectories were straighter and smoother in non-Newtonian solutions as compared with those in Newtonian solutions. We also observe smoother swimming trajectories in our viscoelastic solutions (Fig.~\ref{fig:fig2_Traject}), and quantify this by computing the average curvature (Materials and Methods, Sec. F) of the cell trajectories as a function the of bulk viscosity (Fig.~\ref{fig:fig5_Curvature}). 

In Newtonian solutions, the curvature remains roughly constant over a range of viscosity increasing from $0.98$ cP to $10.5$ cP.  In contrast, the average trajectory curvature in the non-Newtonian solutions decreases as the bulk viscosity rises. A likely reason for the reduction in precession (wobble) has been previously explained \cite{patteson2015running} to be the role of shear induced normal stresses generated by the rotating cell body, an explanation that remains appealing.  However, although the trajectories indeed become straighter, we believe that this phenomenon plays only a subtle role in changing the swimming speed. To estimate the effect of cell body precession on swimming speed, a modified RFT (SI) given by Darton \emph{et al.} \cite{darnton2007torque}, is used to estimate the swimming speed subject to different wobbling angles, $\phi$. Assuming a constant torque motor, the calculated swimming speed increases only about 10\% as $\phi$ changes from $0$ to $\pi /2$ (Fig. SI1) - far less than the observed changes in swimming speed. Furthermore, despite this analysis, and as mentioned earlier, it is not clear that the cell precession reduces swimming speed.  Both Liu et al. \cite{liu2014helical} and Constantino et al. \cite{constantino2016helical} argued that such motion may, under some conditions, enhance the swimming efficiency of bacteria. For these reasons, we argue that the change in cell precession due to shear-induced normal stress, although present, is likely insufficient to explain the speed enhancement observed (Fig.~\ref{fig:fig1_speed}).

\subsection{Shear-thinning enhances swimming speed.}
Elasticity of polymer solution has been shown to  enhance the speed of helical swimmers over a range of Deborah numbers, $De$ $\sim$ $0$ $-$ $2$ \cite{spagnolie2013locomotion,liu2011force} (the Deborah number compares the flagellar rotation rate with the characteristic  relaxation time, $\tau$, of the fluid), and the highest enhancement happens at $De$ $\sim$ $0.7$. Our estimated $De$ number remains in the range of $0.01$ $-$ $0.50$ according to the measured relaxation time $\tau$ (Table.~\ref{table:table2}) and calculated flagellar rotation rate $\omega_f$, which lies in the enhancement region. However, the results from both Spagnolie \emph{et al.}'s numerical study \cite{spagnolie2013locomotion} and Liu \emph{et al.}'s \cite{liu2011force} experiments show that the largest increase in swimming speed is less than $20$\% of the speed achieved in a Newtonian solution with same viscosity. The significant speed enhancement observed in the present experiment seems to be too high to be explained solely by the viscoelastic behavior of the non-Newtonian medium.

In addition to viscoelastic effects, the effect of shear-thinning behavior \cite{katona2013rheological}, has also been proposed to explain the speed increase of flagellated bacteria swimming in polymer solutions \cite{martinez2014flagellated,gomez2017helical,zhang2018reduced}. To preserve a torque free system the cell body rotation rate, $\omega_{c}$, is much smaller than the flagella rotation rate, $\omega_{f},$ and the shear rate, $\dot{\gamma}$, near the flagella, which is estimated as $\dot{\gamma}_{f} = \omega_{f}R/r_{0}$ \cite{martinez2014flagellated}, reaches as high as $10^{4}$ $s^{-1}$. Here $R$ and $r_{0}$ are the radius of flagellar bundle and filament respectively. In contrast, due to its lower rotation speed and larger size, the shear rate near cell body remains much lower: $\dot{\gamma}_{c}$ $\sim$ $10^{2}$ $s^{-1}$. Adopting the modified RFT proposed by Martinez \emph{et al.} \cite{martinez2014flagellated} which assumes different viscosities for the flow around the cell and the flagella, we have theoretically calculated the swimming speed using the measured motor torque (Fig~\ref{fig:fig4_Motor}), shear-thinning behavior of the non-Newtonian solutions (Table.~\ref{table1}) and assuming (i) a cell shear rate of $200$ s$^{-1}$ and (ii) a flagella shear rate ranging between $4000$ and $7000$ s$^{-1}$. The result (Fig.~\ref{fig:fig1_speed}, blue markers) shows a very good agreement with the experimental observations, with the range of shear rates bracketing the measured swimming speeds.  Shear-thinning thus seems to have a much stronger effect on swimming speed than viscoelastic effects have through cell precession or flagellar propulsive efficiency.

\subsection{Shear-induced normal stress reduces flagellar bundling time}

Even if the predominant influence on swimming speed appears to be shear thinning, shear indcued normal stress nevertheless plays a role in cell motility. Here we demonstrate that this non-Newtonian phenomenon affects the swimming behavior and the bundling mechanics for wild-type cells that exhibit run-and-tumble behavior.  Qu \etal argue \cite{qu2018changes} that the flagellar bundling time is extended with increased viscosity in Newtonian solutions and they demonstrate that an increased skewness in the distribution of swimming speeds reflects the change of bundling time in viscous media. It is equally interesting to understand how non-Newtonian effects affect the flagellar bundling process for wild-type cells. As shown in Fig.~\ref{fig:fig3_WT}(a), changes in the average skewness of wild-type cell speed distribution as the Methocel concentration rises suggest that the bundling time of \emph{E.Coli} cells in Methocel is initially increasing with respect to viscosity but then decreasing at higher polymer concentrations. 

Understanding the mechanics of the bundling process is necessary to explain this phenomenon. It has been experimentally established that the bundling process is a purely hydrodynamic process in Newtonian solutions \cite{kim2003macroscopic}. More recently, Man \emph{et al.} \cite{man2017bundling} estimated the hydrodynamic interactions between rotating adjacent elastic rods and clarified the force balance during the bundling process. In Newtonian solutions, the hydrodynamic interactions are balanced by the viscous drag and the bending rigidity (elastic force) of the flagellar filaments. Since we are in the low $Re$ number regime, the force balance on each filament is written as
\begin{equation}
    f_{e} + f_{h} + f_{v} = 0,
\end{equation}
where $f_{e}$, $f_{v}$ and $f_{h}$ refer respectively to elastic and viscous stresses, and hydrodynamic interaction acting on the filament. Two dimensionless numbers are used to describe the relations between these three forces. The ``Sperm number'', $Sp$, quantifies the balance between viscous drag and elastic force \cite{Machin1963,kim2003macroscopic}, and is defined as 
\begin{equation}
    SP = (\frac{\xi_{\perp}\omega_{f}L^{4}}{EI})^{1/4},
\end{equation}
where $\xi_{\perp}$ is the viscous drag coefficient of a slender body on perpendicular direction \cite{lighthill1976flagellar} defined as
\begin{equation}
    \xi_{\perp} = \frac{4\pi \mu}{log(L/r_{0})}.
\end{equation}
$EI$ is the bending modulus of the filaments \cite{landau1986theory} and $L$ is the length of the flagellar filament. The typical value of $Sp$ number of \ecoli is on the order of $1$ \cite{man2017bundling}, indicating the viscous and elastic stress are on the same order of magnitude. The ``Bundling number'', $Bu$, compares the driving force (hydrodynamic interaction among the filaments) in the bundling process to the viscous force \cite{man2017bundling} and is defined as
\begin{equation}
    Bu = \frac{r_{0}^{2}Sp^{4}}{c^{2}},
\end{equation}
where $c$ is the separation of the filament. 
The range of $Bu$ number (with $\omega_{f}$ $\sim$ $100$ Hz) lies in $0.1$ $\sim$ $1$ confirming, not surprisingly, that there  exists a balance between the viscous and bundling forces during the flagellar bundling process of \ecoli.

Since we have argued that shear-induced normal stress plays a role in reducing the cell precession, resulting in straighter swimming trajectories, we also suspect the changes in the skewness of the speed distribution, and the bundling dynamics might also be due to shear-induced normal stress acting on the flagellar filaments. For bundling in a non-Newtonian system, the force balance is rewritten schematically as
\begin{equation}
    f_{e} + f_{h} + f_{v} + f_{n}= 0,
\end{equation}
where we have added $f_{n}$ as the shear-induced normal stress. We can estimate $f_n$ by assuming that we can represent the elasticity of the fluid with a single relaxation time and using an Oldroyd-B model \cite{oldroyd1950formulation} to estimate the forces on a rod of radius $r_o$ rotating at a fixed frequency, $\omega_{f}$, in a large cylindrical container of radius $R_{0}$. Assuming the form of the fluid velocity 
\begin{equation}
    \mathbf{u} = v(r)\bm{\hat{\theta}},
\end{equation}
the rate of strain tensor, $A$, is then given by
\begin{equation}
    A = (\frac{\partial v}{\partial r}-\frac{v}{r})(\bm{\hat{\theta}} \bm{\hat{r}} + \bm{\hat{r}} \bm{\hat{\theta}}).
\end{equation}
The total viscosity of the solution is written as $\mu$ = $\mu_{s}$ + $\mu_{p}$ \cite{pak2015theoretical} where $\mu_{s}$ and $\mu_{p}$ are the solvent and polymer viscosities respectively. The stress tensor, $S$, in a polymer solution can be written as
\begin{equation}
    S = \mu_{S}A + S_{p},
\end{equation}
where $S_{p}$ is the stress due to the polymer contribution. Inserting this into the governing equation for an Oldroyd-B model \cite{oldroyd1950formulation} we find that
\begin{equation}
    S + \tau \overset{\nabla}{S} = \mu (A + \frac{\mu_s}{\mu}\tau \overset{\nabla}{A} ).
\end{equation}
Using this with the momentum balance:
\begin{equation}
    \nabla p = \nabla \cdot S , 
\end{equation}
and the continuity equation: 
\begin{equation}
    \nabla \cdot \bm{u} =0,
\end{equation}
we can show that the velocity field is given by 
\begin{equation}
    v(r) = r_{0}^{2} \omega_{f} \frac{R_{0}^{2}-r^{2}}{r(R_{0}^{2}-r_{0}^{2})} , 
\end{equation}
and the pressure field by
\begin{equation}
    p = 2\mu \tau(1-\frac{\mu_{s}}{\mu}) \omega_{f}^{2} \frac{r_{0}^{4}R_{0}^{4}}{r^{4}(R_{0}^{2}-r_{0}^{2})^{2}}.
\end{equation}
The torque per unit length is then given by
\begin{equation}
    \int_{0}^{2 \pi} r_{0} S_{r\theta} d\theta = -4\pi \mu r_{0} \omega_{f}
\end{equation}
and the normal stress is given by
\begin{equation}
    f_{n} = 2\mu \tau(\frac{\mu_{s}}{\mu}-1) \omega_{f}^{2} \frac{R_{0}^{4}}{(R_{0}^{2}-r_{0}^{2})^{2}}.
\end{equation}
In the case of $R_{0}$ $\rightarrow$ $\infty$, $f_{n}$ is written as
\begin{equation}
    f_{n} = 2 \tau (\mu_{s} - \mu) \omega_{f}^{2}.
\end{equation}
The shear-induced normal stress acts like a ``strangulation'' around the filament. For a multi-filament system, which is the case of a wild type \ecoli during tumbling, the strangulation in all directions forces all filaments to come together. Taking a two filaments (two slender rods separate parallelly) case for example, if each one experience strangulation stress, the total force on this system is pointing inwards to the center. As a result, this helps the flagella bundle faster during a tumble event.

To estimate such effect quantitatively, We define a dimensionless number, the ``Strangulation'' number ($Str$), which compares the shear-induced normal stress to the viscous stress during the bundling process:
\begin{equation}
    Str = \frac{2 \tau (\mu - \mu_{s}) \omega_{f}^{2}}{\xi_{\perp}\omega_{f}} = \frac{2 \tau (\mu - \mu_{s}) \omega_{f} }{\xi_{\perp} } = \frac{2 De (\mu - \mu_{s}) }{\xi_{\perp}}.
  \label{equation:equation5_32}
\end{equation}
For the current study, the $De$ number of the low-concentration non-Newtonian solutions were measured to be between $0.01$ and $0.5$ (Table.~\ref{table:table2}), leading to Strangulation numbers that range between $0.01$ at the lowest concentrations to $0.39$ at the higher concentrations (using viscosity data at low shear rate in Materials and Methods Section E and flagellum geometry in Table. SI1). At low concentrations, the normal stresses can be neglected, and the observed rising skewness of the speed distribution (Fig.~\ref{fig:fig3_WT}(a)) is similar to that observed in the Newtonian fluid, reflecting the longer bundling time driven by the increased bulk viscosity. However in solutions with higher concentration, the  Strangulation number is non-negligible, indicating that the shear-induced normal stresses will play a part in the flagellar bundling process. Since the strangulation force tends to push the filaments together, the normal stresses act to reduce the bundling time and this explains the observed drop in the skewness at high polymer concentrations (Fig.~\ref{fig:fig3_WT}(a)). However, since the non-Newtonian solution we used is also shear thinning, the actual viscosity experienced by the flagella could be much smaller. Using the viscosity measured at $\dot{\gamma}$ = $10^{4}$ $s^{-1}$ (Materials and Methods Section E), the estimated Strangulation numbers range from $0.01$ to $0.36$, which shows no significant difference compare to the results estimated using viscosity measured at low shear rate. Actually, from Equ.$17$ we know that $Str$ is independent of viscosity $\mu$ if $\mu$ $\gg$ $\mu_s$. Even if the shear thinning behavior of solution at high polymer concentration is quite strong, the viscosity at high shear rate is still much greater than the solvent viscosity $\mu_s$.

\section{Conclusions}
The two principal effects of non-Newtonian fluids - shear thinning and viscoelasticity - have long been suspected of affecting the speed and character of swimming flagellated bacteria. However, separating these effects has been complicated by both the run-tumble behavior of wild-type multi-flagellated cells as well as complex behavior associated with the polymeric solutions on the level of cell activity. Although we were unable to clearly separate these two effects either in our study because finding a material which is harmless to cell and exhibits only one non-Newtonian behavior is difficult, our experimental results done with smooth swimmers suggest shear thinning is the dominant factor on the speed enhancement of \ecoli in non-Newtonian fluid. The two-viscosity model \cite{martinez2014flagellated} (high viscosity near cell body and low viscosity near flagellum due to different shear rate) explains our observed swimming speed enhancement well. Viscoelastic effects, both in reducing the cell precession and in increasing the propulsive effectiveness contribute to faster swimming speeds as well, but we argue that they are of lesser importance. In addition, from the results and analysis of wild-type \ecoli swimming in non-Newtonian fluid, we show that shear-induced normal stress does change the run-tumble process and specifically can shorten the bundling time as reflected by the change in speed distribution.

Although these results contribute to a better understanding of swimming in non-Newtonian solutions, further work is needed. It is important to separate the non-Newtonian effects experimentally to fully understand the role of each factor on the speed enhancement of flagellated bacteria. For those looking at the complexities of multi-flagellar motions, bundling, unbundling, experiments in non-Newtonian solutions visualizing the flagellar filaments \cite{turner2016visualizing,turner2000real} are particularly necessary. All the theoretical work done in this study are still viscous analysis since the solutions we used are weakly elastic and the two viscosity model has been proposed before \cite{martinez2014flagellated,magariyama2002mathematical} for solving swimming problem in a shear thinning fluid. Nevertheless, a detailed theoretical study on how non-Newtonian effect, especially elasticity, changes bacterial swimming behavior is crucial in this area.  

\section{Materials and Methods}

\subsection{Cell preparation}
The cells used in the experiments were smooth swimming \emph{E.coli} (Strain: K12 HCB1736) and wild type \emph{E.coli} (Strain: K12 AW405). The wild type cell is known to have a ``run and tumble" motility \cite{berg1972chemotaxis} while the smooth swimming cell does not tumble. The culturing procedure for both strains was identical. A single colony was selected from an  agar plate and cultured in $10$ ml of T-Broth ($1$ L  water, $10$ g Tryptone and $5$ g  NaCl) by rotating at $200$ rpm (Southwest Science, Incu-Shaker Mini) for $16$ h at $30^{\circ}$C. $20$ $\mu$l of bacteria suspension was cultured again in 10 ml of T-Broth for $4$ h until the mid-exponential growing phase of \emph{E.coli}. The bacterial suspension was washed three times by centrifuging at $2000$ rpm (Eppendorf, MiniSpin Plus) for $8$ minutes and re-suspending in fresh motility buffer ($1$ L of water, $11.2$ g $K_{2}HPO_{4}$, $4.8$ g $KH_{2}PO_{4}$, $0.029$ g EDTA, $3.9$ g NaCl; pH $7$-$7.5$). The final suspension was diluted three-fold before conducting experiments.

\subsection{Polymer solutions}
Ficoll 400 and Methocel 90 HG were used to produce Newtonian and non-Newtonian polymer solutions respectively. A 15\% (wt/vol) stock solution of Ficoll 400 (Sigma-Aldrich) and a 0.5\% (wt/vol) stock solution of Methocel 90 HG (Sigma-Aldrich) was prepared by dissolving the polymer in deionized water and rotating overnight at 200 rpm (Southwest Science, Incu-Shaker Mini). The polymer solution was dialyzed for 1 week (Spectra/Por 2 Dialysis Trial Kit; 12$–$14 kD MWCO, 23 mm flat-width membrane) to remove short chain polymer fragments. The final polymer concentration was calculated by measuring the weight before and after evaporating the solvent for $6$ h at $60^{\circ}$C and placing the solution for $4$ hours in vacuum until the weight reached a constant value. 

\subsection{Test fixture}
Cell motility was observed by placing a small volume of the cell suspension into a test fixture consisting of a ``swimming pool'' cut from a 1.5-mm thick film of polydimethylsiloxane (PDMS) and sandwiched between a No. 1 glass slide and a No. 1.5 glass cover slide.

\subsection{Real-Time 3D Digital Tracking Microscopy}
A 3D digital tracking microscope was used to observe the swimming behavior of the cells. The system was identical to that described by Qu \textit{et al.} \cite{qu2018changes}. The cells are observed using a Nikon TE200 inverted microscope with a CFI Plan Fluor20XMI objective and PCO edge 5.5 sCMOS camera. A 2D translational stage (Prior) was used to move the fixture in the $x-y$ plane,  parallel to the focal plane. A computer-controlled piezo objective holder (Physik Instrumente, PI P-725.4CL) was used to rapidly change the location of the focal plane. A  $320 \times 240$ pixel image was acquired at 80 fps, and a real-time algorithm, written in C++ and OpenCV detected the position (centroid) of a single cell in the image and moved the stage and objective to maintain the cell in focus and within the field of view. The system was able to track the position of a motile cell with $1$ $\mu$m precision.

\subsection{The rheological behavior of polymer solutions}\label{Rheology_section}
\begin{figure*}[ht]
\centering
\includegraphics[width=16cm]{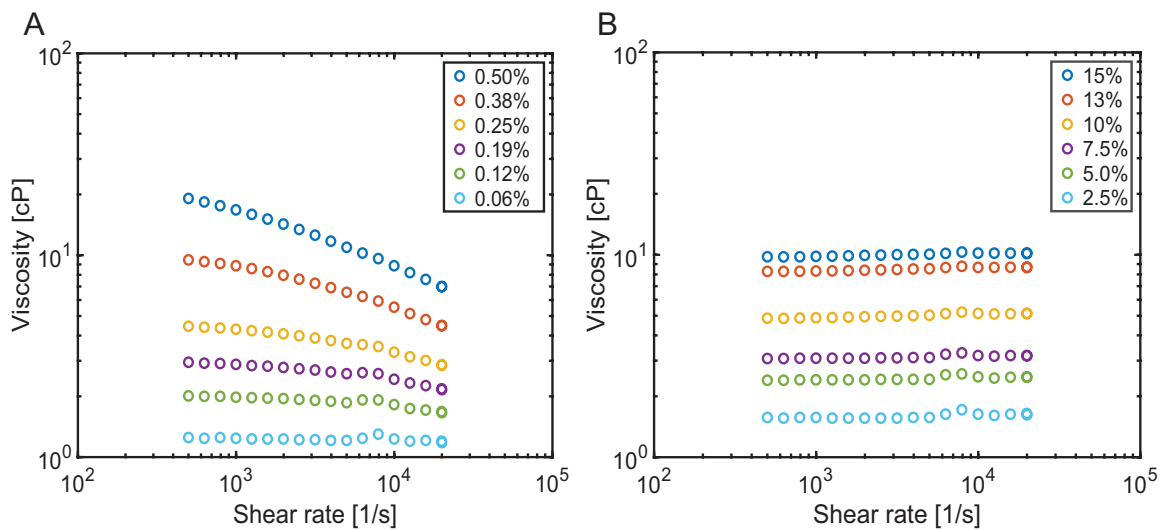}
\caption[Shear viscosity of polymer solutions.]{Shear viscosity of polymer solutions. A, Methocel solutions. B, Ficoll solutions.}
\label{fig:fig6_Rheology}
\end{figure*}

To quantitatively understand the rheological behavior of the polymer solutions, a cone-and-plate rheometer (TA instrument, AR 2000) was used to measure the steady shear rheology of both Methocel and Ficoll solutions at various shear rates, ranging from $500$ $s^{-1}$ to $20000$ $s^{-1}$, using $40$ mm, $0.5^{\circ}$ cone. The shear dependent viscosity, shown in Fig.~\ref{fig:fig6_Rheology}, demonstrates that Methocel solution exhibits strong shear-thinning at high concentrations, while the viscosity of Ficoll solution is nearly shear independent. A non-linear curve fitting using a power-law model \cite{barnes1989introduction} $\mu = m \dot{\gamma}^{n-1}$  is applied to the shear viscosity measurements of the Methocel solutions. For an ideal Newtonian solutions, the power law index $n$ is $1$, while for the Methocel solutions, the shear-thinning index ranges from $0.989$ at low concentration to $0.736$ at the highest concentration tested (Table.~\ref{table1}).

\begin{table}[ht]
\begin{ruledtabular}
\centering
\caption{Consistency index, $m$,  and exponent, $n$, of Methocel solutions using power-law model.}
\begin{tabular}{ccccccc} 
\hline
Conc. [\%]  & 0.063 & 0.125 & 0.188 & 0.250 & 0.375 & 0.500 \\ 
\hline     

 $m$ & 0.001 & 0.003 & 0.005 & 0.009 & 0.034 & 0.103  \\
 $n$ & 0.989 & 0.955 & 0.923 & 0.885 & 0.803 & 0.736   \\

\end{tabular}
\label{table1}
\end{ruledtabular}
\end{table}

The relaxation times of Methocel solution at different concentrations were measured previously using high-speed single particle microrheology \cite{mason1997particle,mason2000estimating,Qu:PhDThesis}. We tracked  the dispersion of nanometer-scale particles in Methocel and Ficoll solutions at various concentrations and measured the displacement of all particles from frame to frame. The mean-squared displacement (MSD) of the particles as a function of time is calculated using Statistical Particle Tracking Velocimetry \cite{guasto2006statistical,Qu:PhDThesis}. The MSD increases linearly over time in Newtonian solutions \cite{miller1924stokes} and this was observed in Ficoll solutions, indicating no viscoelastic behavior. For the measurement done in Methocel solutions, a nonlinear relation between MSD and time is observed \cite{Qu:PhDThesis}. Then the viscoelastic spectrum $G(s)$ and relaxation time $\tau$ is calculated using the method given by Manson et al. \cite{mason1997particle}. Results are given in Table.~\ref{table:table2}. We also calculated the Deborah number, $De$ = $\tau \times \omega_f$ and the results are included in Table.~\ref{table:table2}. The flagellar rotation rates are computed with modified RFT (the two viscosity model) at different polymer concentrations.

Note the viscoelastic behavior measured using particle dispersion is at almost zero shear rate. It is true that the relaxation time could be different given the actual experimental condition. However, for materials following linear viscoelastic models such as the Kelvin-Voight model, the relaxation time depends solely on the viscosity and Young’s modulus under constant applied stress \cite{oswald2009rheophysics} and a smooth swimming cell exert a roughly unchanged stress to the fluid since its speed (flagellar rotation rate) is nearly constant. It has also been shown experimentally that the relaxation time is measured to be constant under different strains and  can be assumed constant for bacteria swimming in viscoelastic fluid \cite{patteson2015running,koser2013measuring}. Thus, we used the relaxation time measured with microrheology to estimate the $De$ number.

\begin{table}
\begin{ruledtabular}
\centering
\caption{Relaxation time of Methocel solutions at various concentrations (data from \cite{Qu:PhDThesis}).}
\begin{tabular}{ccc} 
                Concentration & $\tau$ [ms] & $De$   \\ 
\hline     
                0.063 \% &      0.76   & 0.036\\
                0.125 \% &      1.88   & 0.080\\ 
                0.250 \% &      2.68   & 0.126\\
                0.500 \% &      9.09   & 0.466\\

\end{tabular}
\label{table:table2}
\end{ruledtabular}
\end{table}

\subsection{Average curvature of 3D swimming trajectory}
The curvature of the swimming trajectory is measured and used to quantify the overall wobbling effect. For an object moving in a 3D space, its position and curvature can be simply described as $\bm{r}(t)$ and $\kappa (t)$ respectively:
\begin{equation}
\kappa (t) = \frac{|\bm{r}'(t) \times \bm{r}''(t)|}{|\bm{r}'(t)^{3}|}.
\label{equation:equation5_2}
\end{equation}
To calculate the curvature from the cell trajectory, we fit a third-order polynomial to $j$ measured bacterial positions. The first and second derivatives of $\bm{r}(t)$ is evaluated from the fitted polynomial, and a local curvature is calculated using Equation.~\ref{equation:equation5_2}. Then a moving window with a time step $\delta t = 1/80$, where $80$ is the frame rate used in the experiment, is applied and in this way the local curvature at different time (location) is estimated.

Averaging over all time gives an measure of the trajectory curvature. For the trajectory in polymer solutions, the number of data points ($j$) for local curvature estimation was chosen depending on the average swimming speed. More data points were chosen for slower swimming cells so as to ensure a similar length was used for estimating local curvature. The number of time intervals ($j-1$) were chosen to be inversely proportional to the average swimming speed. We choose $j = 7$ for mean speed $v = 25$ $\mu$m/s.

\begin{acknowledgments}
We are grateful to Coli Genetic Stock Center (Yale University) and to Howard Berg for bacterial strains and advice. We thank Anubhav Tripathi for the help in measuring the fluid viscosities.  A special thanks is due to Saverio Spagnolie who provided considerable insight to the modeling of the filament in a viscoelastic fluid and who suggested the name ``Strangulation number''.  This work was supported by the National Science Foundation (CBET 1336638).
\end{acknowledgments}

\bibliography{apssamp}

\end{document}